\documentclass[12pt]{iopart}

\usepackage[pdftex]{graphicx}
\usepackage{epstopdf}

\begin{document}

\title{Electrical control of spontaneous emission and strong coupling for a single quantum dot}

\author{A Laucht$^{1}$, F Hofbauer$^{1}$, N Hauke$^{1}$, J Angele$^{1}$, S Stobbe$^{2}$, M Kaniber$^{1}$, G B\"{o}hm$^{1}$, P Lodahl$^{2}$, M-C Amann$^1$ and J J Finley$^1$}

\address{$^1$ Walter Schottky Institut, Technische Universit\"at M\"unchen, Am Coulombwall 3, D-85748 Garching, Germany}
\address{$^2$ DTU Fotonik, Department of Photonics Engineering, Technical University of Denmark, DTU - Building 345V, DK-2800 Kgs. Lyngby, Denmark}
\ead{finley@wsi.tum.de}

\begin{abstract}
We report the design, fabrication and optical investigation of electrically tunable single quantum dot - photonic crystal defect nanocavities operating in both the weak and strong coupling regimes of the light matter interaction. Unlike previous studies where the dot-cavity spectral detuning was varied by changing the lattice temperature, or by the adsorption of inert-gases at low temperatures, we demonstrate that the quantum confined Stark effect can be employed to quickly and reversibly switch the dot-cavity coupling simply by varying a gate voltage. Our results show that exciton transitions from individual dots can be tuned by $\sim4$ meV relative to the nanocavity mode before the emission quenches due to carrier tunneling escape. This range is much larger than the typical linewidth of the high-Q cavity modes ($\sim100 \mu$eV) allowing us to explore and contrast regimes where the dots couple to the cavity or decay by spontaneous emission into the 2D photonic bandgap. In the weak coupling regime, we show that the dot spontaneous emission rate can be tuned using a gate voltage, with Purcell factors $\geq7$. New information is obtained on the nature of the dot-cavity coupling in the weak coupling regime and electrical control of zero dimensional polaritons is demonstrated for the highest-Q cavities ($Q\geq12000$). Vacuum Rabi splittings up to $\sim130\mu$eV are observed, much larger than the linewidths of either the decoupled exciton ($\gamma\leq40\mu$eV) or cavity mode. These observations represent a voltage switchable optical non-linearity at the single photon level, paving the way towards \emph{on-chip} dot based nano-photonic devices that can be integrated with passive optical components.
\end{abstract}
\submitto{\NJP}
\maketitle
Spontaneous emission is very fundamental and is often regarded as an inherent property of an excited atom, molecule or quantum dot. However, this view overlooks the fact that it is not only a property of the emitter, but of the combined emitter - vacuum system\cite{har89}. The irreversibility of spontaneous emission comes about due to the infinite number of vacuum states available to the emitted photon. If the photonic environment is modified, for instance by placing the emitter within a cavity, spontaneous emission can be inhibited, enhanced or even made to become reversible\cite{har92,khi06}.\\
For semiconductor quantum dots in high finesse cavities a wide range of novel optoelectronic devices have been realized using such cavity quantum electrodynamic phenomena. Examples include high efficiency single photon sources\cite{eng05,cheng06,kan07,kan08a}, low-threshold, high bandwidth nanocavity lasers\cite{pai99,par04,alt06} and even \emph{single-quantum dot} optical components like mirrors\cite{eng06} and phase shifters\cite{fush08}.\\
All such single-quantum dot cavity quantum electrodynamic devices call for a method to precisely control the spectral detuning between the quantum dots and the cavity mode $(\Delta=\hbar(\omega_{QD}-\omega_{cav}))$. Until now this has been done by slowly tuning the lattice temperature\cite{rei04,yos04,pre07} or by condensing inert-gases at low temperatures\cite{mos05,hen07}. A major drawback of both of these methods is that they are slow, rendering them impractical for future single-quantum dot devices.\\
In this paper we demonstrate an electrically contacted single-quantum dot photonic crystal nanocavity that allows $\Delta$ to be rapidly and reversibly switched over $\sim4$ meV using the quantum confined Stark effect\cite{fin04}.
In the weak-coupling regime of cavity quantum electrodynamics we observe a voltage switchable Purcell effect\cite{pur46}
($F_{p}\geq7$) \emph{as well as} non-resonant coupling\cite{hen07,kan08b} which is active for $\Delta\leq5$ meV. For cavities with $Q\geq10000$ we observe strong coupling\cite{and99,Lau08} with vacuum Rabi splittings up to $\sim 130$ $\mu$eV, representing a voltage switchable optical non-linearity.
Our techniques could be readily extended to realize \emph{on-chip} photonic crystal single-quantum dot optoelectronic devices
with modulation bandwidths in the GHz range.
This may pave the way towards the realization of on-chip, coherent single-quantum dot quantum photonic devices.

The samples investigated are GaAs p-i-n photodiode structures patterned into membrane photonic crystals (PCs) as depicted schematically in Fig. 1a (see Methods)\cite{hof07}. A single layer of InGaAs self-assembled quantum dots (QDs) with an areal density $\leq20$ $\mu$m$^{-2}$ are incorporated in the intrinsic region of the device, such that they are subject to an axial electric field which can be tuned by varying the voltage applied across the p-i-n junction ($V_{app}$).  An array of 5$\times$5 PCs were defined into the photodiode window in the top-contact, at the center of each of which $L3$ nanocavities\cite{aka03} were placed (Fig. 1b).  The $L3$ nanocavities support six strongly localized modes within the 2D-photonic bandgap with Q-factors ranging from $3000\leq Q\leq15000$\cite{aka03}.  As discussed below, this allows measurements to be performed in both the weak and strong coupling regimes of the light-matter interaction. Fig. 1c shows a typical current-voltage trace recorded
\begin{figure}
\centering
\includegraphics[width=12cm]{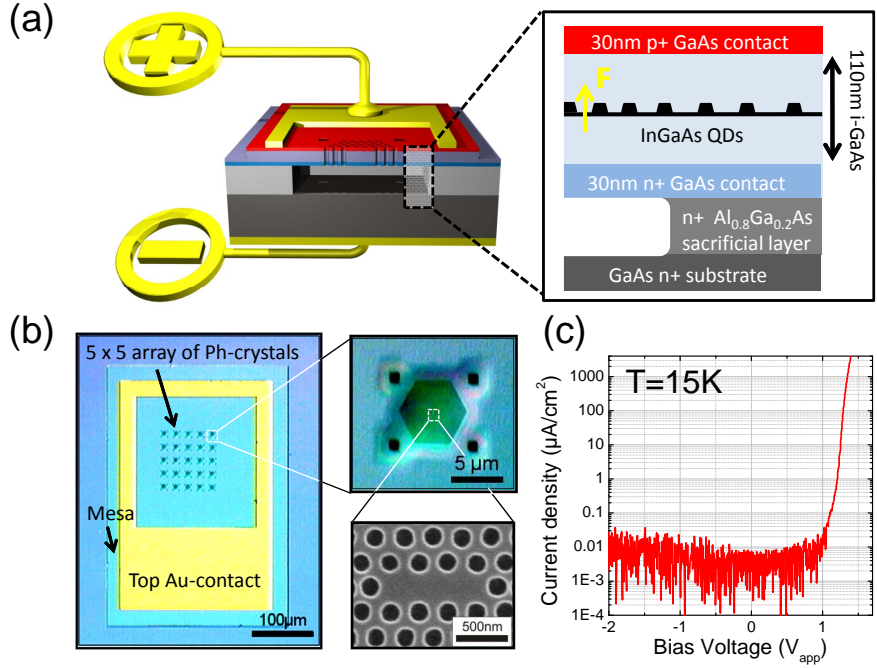}
\caption{\textbf{Electrically tunable single-QD PC nanocavities investigated.}
\textbf{a}, Schematic cross-sectional representation of the device and layer sequence of the active region. The polarity of the static electric field ($F$) is indicated.
\textbf{b}, Microscope image of one of the diode windows showing the 5$\times$5 array of PCs and image of higher magnification depicting a single PC. The scanning electron microscope image shows one of the $L3$ cavites at the center of each PC.
\textbf{c}, Current-voltage characteristic of the device recorded without illumination.
}
\end{figure}
without illumination at $T=15$ K. Clear rectifying behavior is observed with a forward bias current onset of $V_{bi}\sim1.1$ V, corresponding to the built-in potential in the membrane p-i-n diode\cite{foot2}, and negligible current flow ($< 0.01$ $\mu$A/cm$^2$) in reverse bias.  From the built-in potential and the device geometry we estimate the static electric field to be $F\sim(V_{bi}-V_{app})/d$, where $d=110$ nm is the thickness of the intrinsic region and $V_{app}$ is the applied electrostatic potential. Thus, we expect that static electric fields in the range $+100$ kV/cm$\leq F\leq0$ kV/cm can be applied parallel to the QD-growth axis, as depicted in Fig. 1a.

Typical $V_{app}$-dependent micro-photoluminescence ($\mu$PL) measurements recorded from one of the $L3$ cavities are presented in the false colour plot in Fig. 2a. A number of sharp emission lines are observed, all of which shift to lower energy as $V_{app}$ reduces, corresponding to increasing $F$. In addition, a prominent feature is observed at $\hbar\omega_{cav}=1269.0$ meV, the energy of which is unaffected by the bias voltage. We identify this feature as being due to the fundamental cavity mode of the $L3$ defect nanocavity and the shifting peaks as single exciton transitions from different QDs located in the \begin{figure}
\centering
\includegraphics[width=12cm]{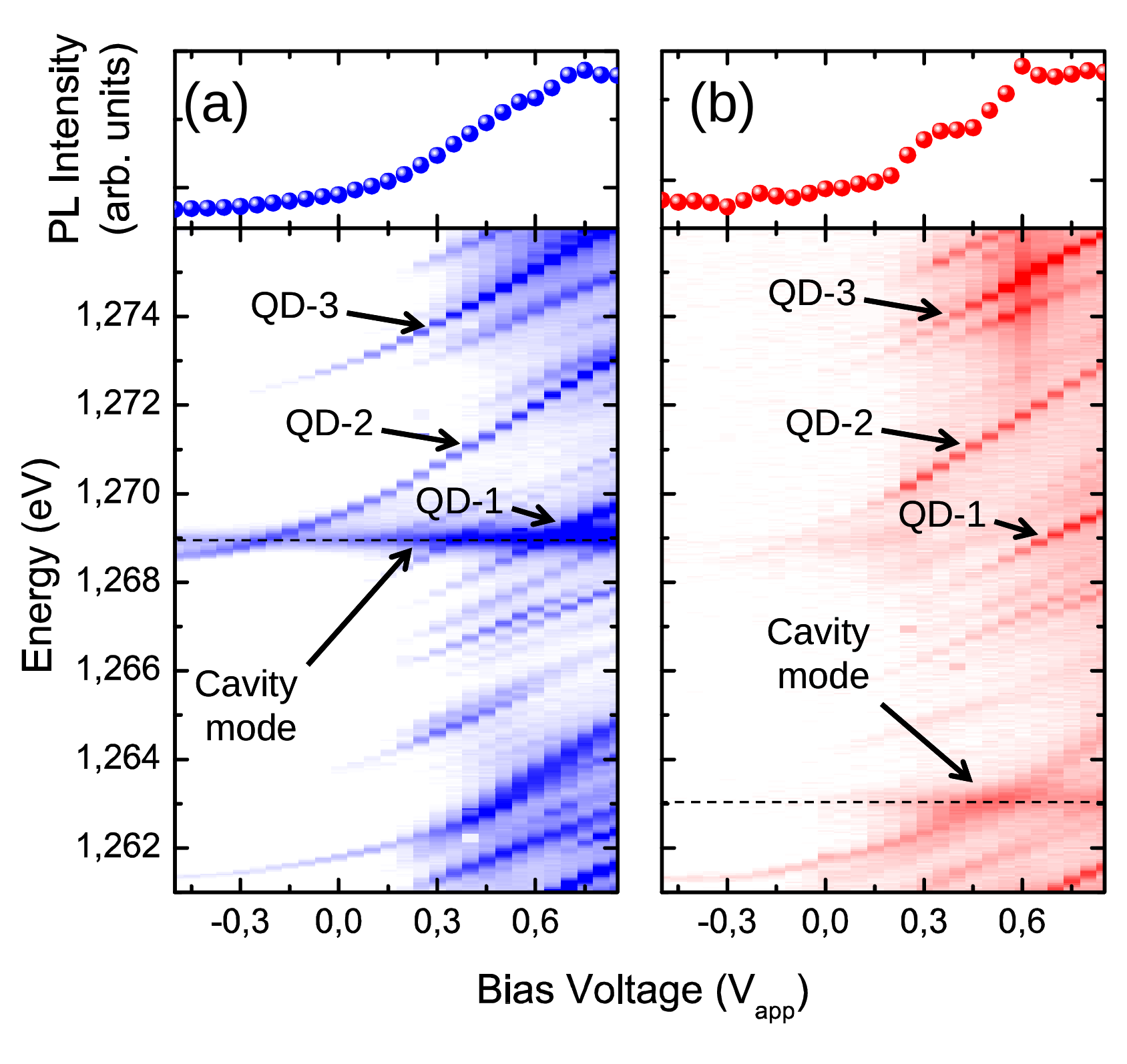}
\caption{\textbf{Micro photoluminescence measurements of a typical single-QD nanocavity at 15 K before and after detuning the cavity mode using nitrogen deposition.}
\textbf{a}, Before nitrogen deposition showing the cavity mode unaffected by the gate voltage, whilst QD transitions shift due to the QCSE. The three QDs discussed in the text are labeled QD-1, QD-2 and QD-3, respectively.
\textbf{b}, Equivalent data to that presented in (a) after adsorption of molecular nitrogen. The upper panels of (a) and (b) show the spectrally integrated $\mu$PL yield as a function of V$_{app}$.
}
\end{figure}
cavities\cite{nom06, kan08c}. The QD peaks shift to lower energy with reducing bias due to the quantum confined Stark effect (QCSE) whilst the mode emission remains fixed due to the negligible variation of the refractive index of the PC GaAs matrix $\sim200$ meV below the direct bandgap. The spectrally integrated luminescence intensity is plotted as a function of $V_{app}$ in the upper panels of Fig. 2. The total $\mu$PL intensity remains approximately constant for $V_{app}\geq0.6$ V and quenches rapidly for $V_{app}<0.6$ V due to carrier tunneling escape from the QDs in the internal field\cite{fin04}. The data presented in Fig. 2a shows that the QD transitions can be tuned by $|\Delta|\sim2-4$ meV relative to the nanocavity mode\cite{foot4} before quenching due to carrier tunneling occurs. Three different QD transitions, labeled \textbf{QD-1}, \textbf{QD-2} and \textbf{QD-3} on Fig. 2a, will be studied using time-resolved spectroscopy below. The tuning range provided by the QCSE is much larger than the linewidth of the high-Q cavity mode ($Q=4020$), allowing us to explore and contrast regimes where QD transitions spectrally couple to the optical cavity, decay by spontaneous emission (SE) into the 2D photonic bandgap or are quenched due to carrier tunneling escape from the QDs.

By controllably adsorbing molecular nitrogen into the PC structure the cavity mode can be controllably red-shifted\cite{mos05,hen07}. Fig. 2b shows $\mu$PL recorded from the same nanocavity after the mode has been red-shifted by $-6$ meV.  We note that after shifting, the emission energies and $V_{app}$-dependence of \textbf{QD-1}, \textbf{QD-2} and \textbf{QD-3} remain unchanged.  Before detuning the mode we systematically performed time-resolved measurements detecting on \textbf{QD-1} and the cavity mode emission as a function of $V_{app}$. Clearly, \textbf{QD-1} is electro-optically fine tuned
through the cavity mode over a bias range where tunneling escape from the QDs is negligible and $\Delta=0$ meV for $V_{app}=+0.65$ V.  Thus, the measured decay dynamics directly reflects on the SE lifetime and its dependence on QD-cavity detuning ($\Delta$). Similar measurements were then repeated on \textbf{QD-1} \emph{after} detuning the mode. These complementary measurements, before and after detuning, permit us to investigate the SE dynamics with and without spectral coupling to the mode, but at equal values of $V_{app}$ and, thus, $F$.

Selected results of our time-resolved measurements are summarized in Fig. 3. Five situations of interest, labeled \textbf{1} - \textbf{5}, are marked on the false color plot in Fig. 3a.  Traces \textbf{1}, \textbf{2} and \textbf{3} in Fig. 3b were recorded with detection on the \emph{cavity mode} with $V_{app}=+0.45$ V, $+0.65$ V and $+0.85$ V, corresponding to \textbf{QD-1} being detuned by $\Delta=-0.7$ meV, $0$ meV and $+0.7$ meV, respectively. In contrast, traces \textbf{4} and \textbf{5} were recorded with detection \emph{on} \textbf{QD-1} for $V_{app}=+0.50$ V and $+0.80$ V, corresponding to $\Delta=-0.6$ meV and $+0.6$ meV, respectively.  When $\Delta\neq0$ the observed decay transients detected on the mode are
mono-exponential and exhibit a characteristic lifetime close to $\tau_{Mode}\sim0.8$ ns\cite{foot5}. In strong contrast, for $\Delta=0$ meV (situation \textbf{2}) emission from \textbf{QD-1} and the cavity mode are superimposed and a clear biexponential decay transient is observed (see Fig. 3b).  The time constants extracted with detection on the cavity mode and \textbf{QD-1} are plotted in the upper and lower panels of Fig. 3c, respectively.  Besides \textbf{QD-1}, two other weaker transitions shift through the cavity mode at $V_{app}=+0.35$ V and $V_{app}=+0.15$ V, respectively.  As for \textbf{QD-1}, each \begin{figure}
\centering
\includegraphics[width=12cm]{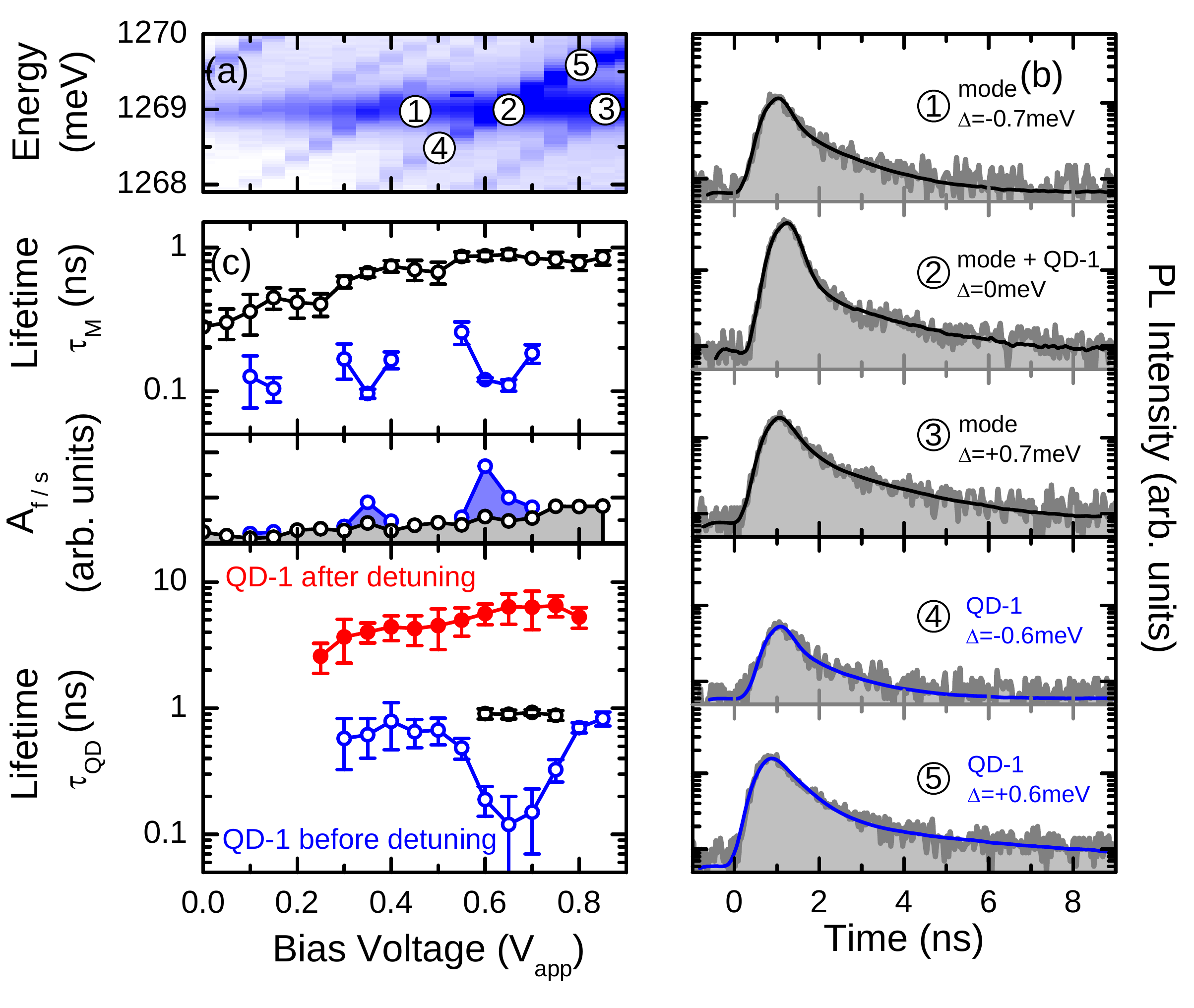}
\caption{\textbf{Lifetime measurements on cavity mode and QD.}
\textbf{a}, False-colour PL plot giving an overview of the investigated situation.
\textbf{b}, Decay transients recorded on the cavity mode (traces \textbf{1, 2, 3}) and a QD (traces \textbf{2, 4, 5}) as indicated in Fig. 3a.
\textbf{c}, (upper panel) Lifetimes measured on the cavity mode as a function of $V_{app}$ (black circles). The blue trace shows the fast component for voltages where a biexponential decay is observed. (middle panel) Corresponding PL intensity for the slow (black) and fast (blue) decay components $A_{f/s}$. (lower panel) Lifetimes measured on \textbf{QD-1} as a function of $V_{app}$ (blue open circles). The black trace shows the slow component of the biexponential decays. The lifetimes recorded on the QD when the cavity mode was shifted away by adsorption of nitrogen are plotted by the red filled circles.
}
\end{figure}
time a transition shifts into resonance with the cavity mode we observe a bi-exponential decay, with the fast component being identified as the Purcell enhanced emission of the QD into the cavity mode\cite{pur46}. This interpretation is confirmed by examining the bias dependent time-integrated amplitudes of the fast ($A_f$) and slow ($A_s$) decay components (Fig. 3c - middle panel). The intensity of the slow decay component (black trace) does not exhibit resonances and decreases monotonically upon reducing $V_{app}$ in a manner that is fully analogous to Fig. 2a (upper panel). In strong contrast, the superimposed fast decay (blue trace) \emph{only} appears when a QD transition is tuned into resonance and is strongest exactly in resonance.

For large QD-cavity detunings (i.e. $|\Delta|\geq\hbar\omega_{cav}/Q\geq315$ $\mu$eV) the QD SE-lifetime remains almost constant at $\tau_{\Delta\neq0}^{QD}\sim0.7$ ns (blue open circles in Fig. 3c - lower panel), whilst for zero detuning it is $\tau_{\Delta=0}^{QD}=0.12$ ns, limited by the temporal response of our detection system\cite{foot3}. This result represents a direct measurement of an electrically tunable Purcell effect, where the ratio of lifetimes yields a Purcell factor of $F_P\geq7$. We note that the dependence of $\tau^{QD}$ on $\Delta$ in Fig 3c (lower panel) reflects precisely the spectral profile of the mode linewidth observed in PL ($315$ $\mu$eV) confirming that the observed SE-rate enhancement is due to the Purcell effect. Close to $\Delta=0$ we detect the cavity mode background emission (black open circles) as a slow component in the decay transients (see e.g. Fig. 3 - situation \textbf{2}).\\
In order to separate the influence of the spectral proximity of the mode from the $V_{app}$-dependence due to tunneling, we repeated the time-resolved measurements on \textbf{QD-1} after strongly detuning the mode via controlled nitrogen deposition\cite{mos05,hen07} (Fig. 2b). The resulting decay transients were found to be purely mono-exponential and the resulting SE-lifetimes are plotted in Fig. 3c (lower panel) by the filled red circles. In strong contrast to the case before detuning, the SE-lifetime reduces monotonically as $V_{app}$ reduces, varying from $\sim6$ ns to $\sim3$ ns as $V_{app}$ is reduced from $+0.7$ V to $+0.3$ V. Resonances are \emph{not} observed as expected since \textbf{QD-1} is now strongly detuned from the cavity mode ($\Delta\sim6$ meV over this bias range).  Although these lifetimes are characteristic of QDs emitting into a 2D photonic bandgap\cite{kan08a}, it is remarkable that they are $\sim8\times$ longer , when compared to the saturation value of $\tau^{QD}_{\Delta\neq0}\sim0.7$ ns before the nitrogen-tuning (compare with open blue circles on Fig. 3c - bottom panel).  This observation indicates that two distinct mechanisms influence the SE-lifetime; (i) Purcell effect due to direct radiative coupling to the cavity mode and (ii) non-resonant coupling between the QD and the cavity mode that is active over a much larger range of detunings. Such a non-resonant QD-cavity coupling mechanism has previously been revealed by photon cross-correlation spectroscopy and clearly plays a role in these experiments\cite{pre07, hen07, kan08b}.

To further elucidate the energy range over which this non-resonant coupling mechanism is active, in Fig. 4 we summarize the $V_{app}$-dependent lifetime measurements performed on \textbf{QD-1}, \textbf{QD-2} and \textbf{QD-3} as a function of energy detuning $\Delta$ from the cavity mode.  Two data sets are plotted for each QD; before (filled blue symbols) and after (open red symbols) detuning from the cavity mode by nitrogen deposition. Each data set represents measurements where $V_{app}$ was varied from $0.8$ V in $-0.05$ V steps, the electric field $F$ increasing as indicated on Fig. 4. The data presented clearly reveal the reduction of the measured SE lifetime at high $F$ due to non-radiative tunneling escape of carriers from the
\begin{figure}
\centering
\includegraphics[width=12cm]{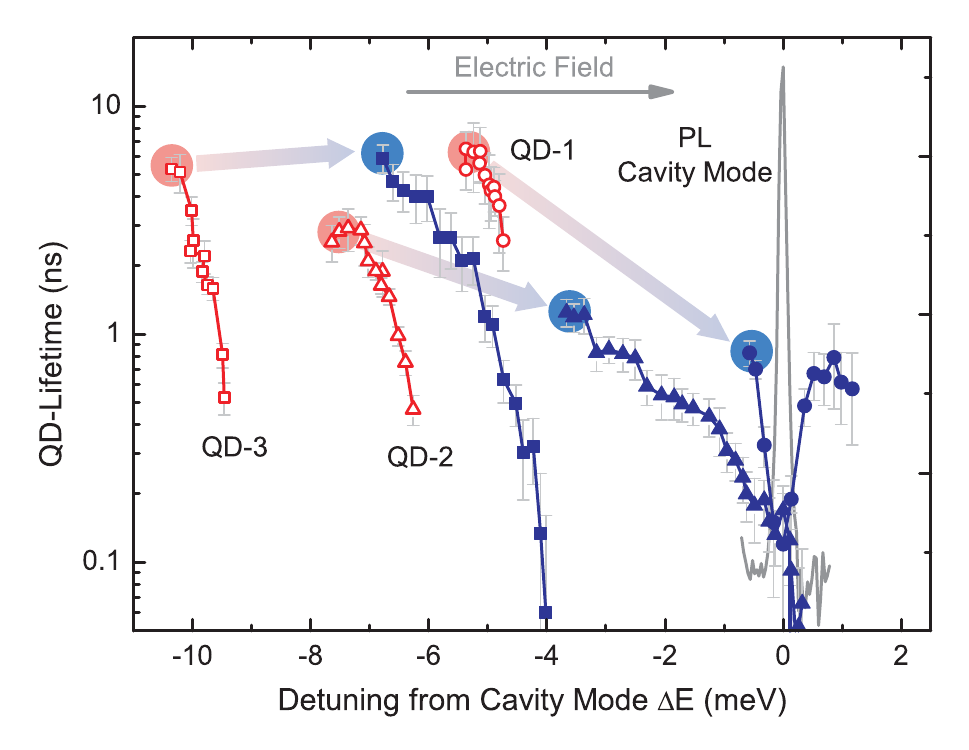}
\caption{\textbf{Comparison of $V_{app}$-dependent lifetime measurements on QD-1, QD-2 and QD-3 plotted versus the detuning from the cavity mode.}
Lifetime-dependent measurements recorded from QD-1 (circles), QD-2 (triangles) and QD-3 (squares) recorded as a function of $V_{app}$.  Filled blue (empty open) symbols correspond to measurements performed on the same QD transitions before (after) detuning the mode by nitrogen deposition. The first point in each data set is recorded at the same value of $V_{app}=+0.8$ V, clearly revealing that the non-resonant coupling mechanism is active for detunings up to $\sim5$ meV as discussed in the text.
}
\end{figure}
QD as discussed in the text.  Comparison of the pairs of data sets from the three QDs reveals information about the energy range over which the non-resonant coupling mechanism is active.  For example, comparison of the first point of the data sets for \textbf{QD-1} (circles), both recorded at $V_{app}\sim+0.8$ V, shows that the lifetime shortens dramatically, before and after shifting, from $\sim6.5$ ns to $\sim0.8$ ns as $\Delta$ reduces from $\sim-5.4$ meV to $\sim-0.5$ meV. For \textbf{QD-2} (triangles) which is detuned by $\Delta=-3.6$ meV at low $F$, a weaker, albeit, similar change of the lifetime is observed from $3.0$ ns to $1.2$ ns.  Only for \textbf{QD-3} (squares), which is strongly detuned from the mode by $\Delta\sim-6.8$ meV for low $F$, is the $V_{app}$ data similar for both cases, before and after detuning. This observation is a strong indication that the non-resonant coupling mechanism is active for detunings as large as $\sim5$ meV, an observation consistent with suggested mechanisms as being due to decay into a continuum of final states for multiply charged QDs\cite{kan08b}.

In the final section of this letter we demonstrate electrical control of zero dimensional polaritons in an high-Q cavity with $Q\sim12500$\cite{foot1}. For a low excitation power of $P\sim1$ Wcm$^{-2}$, the QD is far from being saturated, and we have the possibility to observe vacuum Rabi splitting. In Fig. 5a, we show a series of PL spectra recorded with high spectral resolution for different bias voltages and observe clear QD - cavity anticrossing. This anticrossing unambiguously demonstrates that the QD-cavity system is in the strong coupling regime\cite{and99} and becomes even clearer when we plot the center peak
\begin{figure}
\centering
\includegraphics[width=12cm]{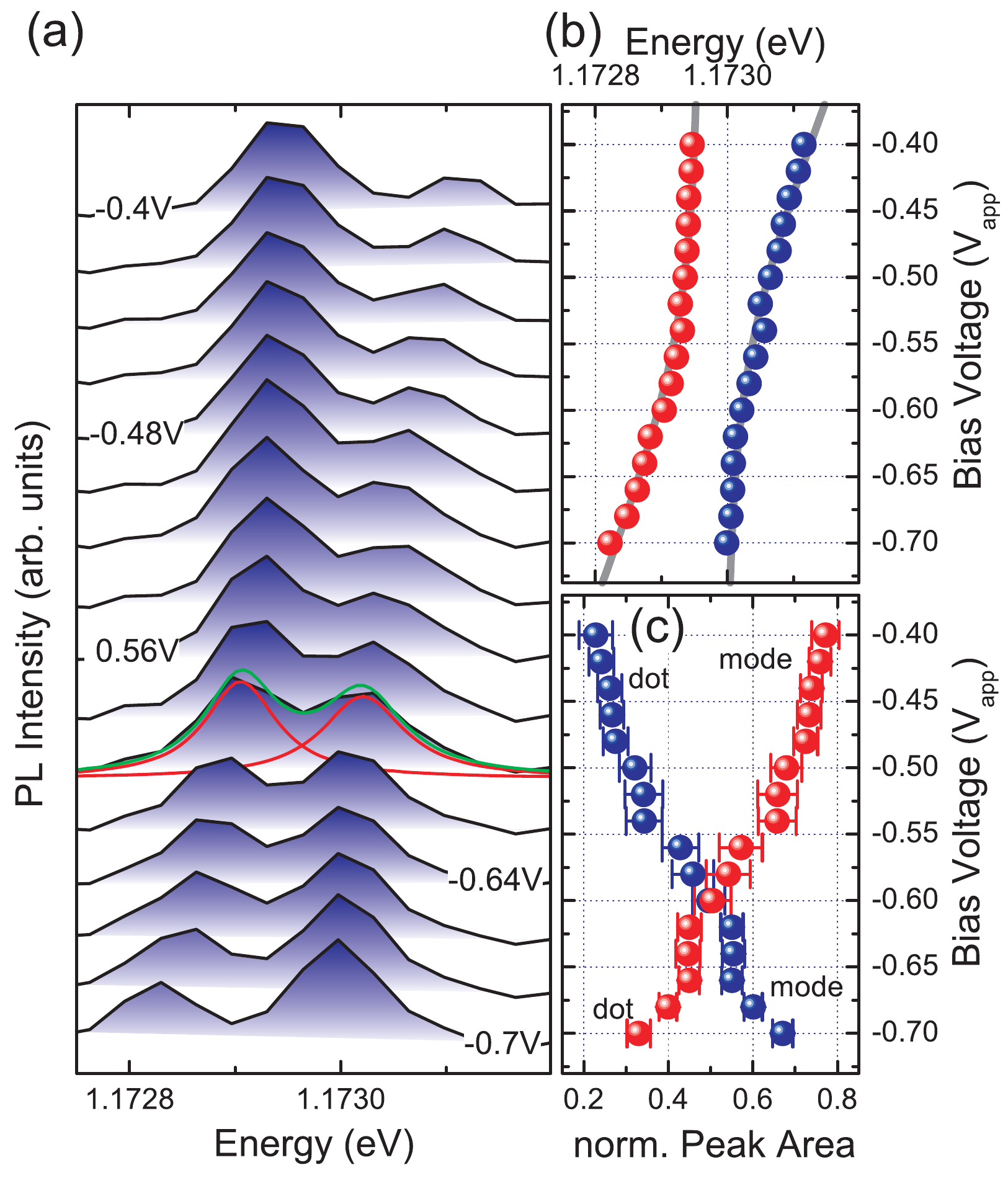}
\caption{\textbf{Demonstration of electrically controlled strong coupling.}
\textbf{a}, Waterfall plot of PL spectra for different bias voltages.
\textbf{b}, Peak centre position for the left (red trace) and the right (blue trace) peak for various detunings. The gray curves show the theoretical expectation fitted to the data.
\textbf{c}, Normalized peak area for the left (red trace) and the right (blue trace) peak.
}
\end{figure}
position of both peaks as a function of applied bias as in Fig. 5b. The gray curves are obtained from theory with parameters fitted to the data and are in excellent agreement with the experiment. From the fits, we extract a vacuum Rabi splitting of $2g=121$ $\mu$eV, similar to values reported\cite{hen07, pre07, yos04, rei04} by other groups for self-assembled QDs. Fig. 5c shows the peak intensities of the higher and lower energy peaks normalized to the total emission intensity plotted for various detunings. For $V_{app}=-0.7$ V, the lower energy peak is predominantly exciton-like and has a weak intensity compared to the higher energy peak that is predominantly cavity-like. As the two peaks are shifted into resonance, the intensities become similar and both peaks represent strongly coupled, single-QD exciton polariton states. When the system is then detuned away from resonance they exchange characteristic properties; the lower energy peak becomes cavity mode-like and the higher energy peak exciton like. These observations unequivocally prove the strong coupling character of this anticrossing in an electrically tunable system.

%Summary
In summary, we have demonstrated electrically tunable single-QD PC nanocavities operating in the weak and strong coupling regimes of the light-matter interaction.  In the weak coupling regime, we demonstrated electrical control of the radiative lifetime, with evidence for a short range reduction of the SE-lifetime due to the Purcell effect and a long range interaction that is active for QD-cavity detunings up to $\Delta\sim5$ meV. For structures operating in the strong coupling regime we have observed an electrically switchable optical non-linearity present at the single-QD, single-photon level.\\

We acknowledge financial support of the Deutsche Forschungsgemeinschaft via the Sonderforschungsbereich 631, Teilprojekt B3 and the German Excellence Initiative via the Nanosystems Initiative Munich (NIM).\\
\newpage
\textbf{Addendum}\\\\
\emph{Semiconductor structure:}
The samples studied were GaAs p-i-n photodiode structures and nominally consist of the following layers: Firstly, we deposited a 500 nm thick, n-type (n=2$\times10^{18}$ cm$^{-3}$, Si-dopant) Al$_{0.8}$Ga$_{0.2}$As sacrificial layer followed by a 35 nm thick, n-type GaAs lower contact layer (n=2$\times10^{18}$ cm$^{-3}$, Si-dopant). This is followed by a 110 nm thick intrinsic GaAs waveguide core into the centre of which we grew a single layer of QDs by depositing 7.36 ML of In$_{0.4}$Ga$_{0.6}$As at 595 $^\circ$C and a rate of 0.04 ML/s. A 35 nm thick p-type (p=2$\times10^{19}$ cm$^{-3}$, C-dopant) GaAs top contact was then grown to complete the structure. After growth we established an Ohmic back contact to the buried n$^{+}$ layers and defined 300 $\mu$m $\times$ 400 $\mu$m photodiode mesas using photolithography and wet etching techniques. 200 $\mu$m $\times$ 200 $\mu$m windows were opened in the metallic top contact for optical access. Inside each window we then formed an array of 5 $\times$ 5 PhC nanocavities by combining electron beam lithography and Cl$_2$-Ar reactive ion etching to define a lattice of cylindrical air holes (radius $r$) arranged in a hexagonal lattice with period a=260 nm and r/a$\sim0.3$. Nanocavities were formed by three missing hole line defects with detuned outer holes. Finally, suspended PhC membrane structures were fabricated by selectively removing the Al$_{0.8}$Ga$_{0.2}$As layer beneath the GaAs waveguide core to leave a free-standing, p-i-n doped GaAs membrane.\\
\\
\emph{Optical characterization:}
The sample was mounted in a liquid He-flow cryostat and cooled down to T=15 K. For excitation, we used a pulsed Ti:sapphire laser (f$_{laser}$=80 MHz, 2 ps duration pulses) laser tuned into resonance with a higher energy cavity mode ($1.305$ eV$<E_{Laser}<1.355$ eV)\cite{nom06, kan08c}. The QD micro-photoluminescence was collected via a 100 $\times$ microscope objective (numerical aperture=0.55) providing a spatial resolution of $\sim$700 nm and the signal was spectrally analyzed by a 0.55 m imaging monochromator and detected with a Si- or an InGaAs-based, liquid nitrogen cooled charge coupled device detector. For time-resolved measurements, we used a fast silicon avalanche photodiode that provided a temporal resolution of $\sim$100 ps after deconvolution.\\


\begin{thebibliography}{1}
\bibitem{foot2} We note that the onset voltage of our photodiodes is typically $\sim1.1$ V instead of the $\sim1.5$ V expected for a GaAs p-i-n structure.  This discrepancy arises due to the low doping level of the GaAs n-contact ($2\times10^{18}$ cm$^{-3}$) and Al$_{0.8}$Ga$_{0.1}$As sacrificial layers.  This low doping level was experimentally found to be required to obtain high finesse cavity modes.
\bibitem{foot5} The measured lifetime for the cavity mode does not reflect the photon loss rate from the cavity, rather the timescale over which the cavity mode is fed via the non-resonant coupling mechanism\cite{kan08b}.
\bibitem{foot3} Since the fast component of the decay is at the resolution limit of our avalanche photo diodes, we can only ascertain that the lifetime is shorter than $\tau_{QD\_on}<0.1$ ns.
\end{thebibliography}
\end{document}